# Unusual Thermally Induced Blueshift and Emission Amplification of $Mn^{2+}$ ions Enable Filter-Free Luminescent Thermal Imaging

Y. Abe[1], M. Szymczak[1,*], J. Zeler[2], L. Marciniak[1,*]

[1] Institute of Low Temperature and Structure Research, Polish Academy of Sciences, Okólna 2, 50-422 Wrocław, Poland

[2] Faculty of Chemistry, University of Wroclaw, 14 F. Joliot-Curie Street, Wroclaw, 50-383, Poland

*corresponding author: l.marciniak@intibs.pl, m.szymczak@intibs.pl

**Abstract**

The shift from point-based thermal sensing to filter-free thermal imaging requires luminescent thermometers that exhibit pronounced and thermally driven spectral changes within spectral regions matching the sensitivity profiles of the R, G and B channels of a digital camera. In this work, we introduce such a system, enabled by the synergistic interplay between (i) thermal redistribution among the vibronic components of the $^4T_1$ excited state of $Mn^{2+}$ ions and (ii) thermally assisted population of this state via optical trap sites. These combined processes result in a simultaneous thermal enhancement and blueshift of the $Mn^{2+}$ emission band associated with the $^4T_1 \rightarrow {}^6A_1$ electronic transition. Consequently, the emission intensity recorded in the G channel increases with temperature, while the luminescence signals detected in the B channel exhibit a corresponding decrease. As demonstrated, $Ca_{19}Zn_2(PO_4)_{14}:Mn^{2+}$, $Ce^{3+}$ supports not only sensitive filter-free thermal imaging, but also two additional ratiometric readout schemes: one based on the intensity ratio of $Ce^{3+}$ and $Mn^{2+}$ emissions, and another based on two distinct spectral regions of the $Mn^{2+}$ emission band, yielding maximum relative sensitivities of 0.42%K$^{-1}$ and 2.7% K$^{-1}$, respectively. This approach introduces a unique thermometric strategy that enable simple, robust, and cost-effective two-dimensional thermal imaging without the need for optical filters or specialized instrumentation.

**Introduction**

Luminescent thermometry, which exploits temperature-dependent variations in the spectroscopic properties of phosphors, has emerged as a powerful alternative to conventional contact thermometry[1–4]. Its growing scientific interest stems from several intrinsic advantages, including high sensitivity, fast response, operational simplicity, and the ability to perform real-time and minimally invasive measurements [5–7]. Importantly, luminescent thermometry enables accurate temperature readout even for moving or rotating objects, substantially reducing measurement time and cost while maintaining high precision[5,8,9]. Although the majority of studies to date have focused on point-based thermal sensing, the full potential of luminescent thermometry is realized in thermal imaging, where spatially resolved temperature maps can be obtained[10–15].

Traditionally, two-dimensional thermal imaging using phosphors is achieved by employing band-pass optical filters matched to the emission bands of the material[10,12,15]. In this method, two images of the object are recorded using different filters, and their pixel-wise ratio yields a two-dimensional temperature distribution. Although conceptually simple and effective, this technique is fundamentally limited by the time required for exchanging filters or mechanically switching them in front of the detector. For systems undergoing rapid or transient thermal processes, this delay becomes prohibitive, effectively preventing accurate thermal imaging.

To circumvent these limitations, a filter-free imaging strategy based on the intrinsic spectral selectivity of the RGB channels of a standard digital camera was recently introduced [11,16]. In this approach, a single photograph provides spatially resolved intensity maps for the red (R), green (G), and blue (B) channels. After appropriate calibration, pixel-wise ratios of these channels allow fast, low-cost, and real-time thermal imaging without the need for tunable filters or dedicated optical components[16,17]. Numerous reports have demonstrated the viability of this concept. However, its successful implementation requires luminescent materials whose emission bands lie within the spectral response windows of the RGB channels and exhibit sufficiently distinct and thermally responsive behaviors. So far, most reported RGB-based luminescent thermometers rely on differences in the thermal quenching rates of independent emission bands[16,17]. Consequently, the achievable sensitivity is often limited by the rate of thermal signal quenching in at least one of the channels. From this perspective, phosphors co-doped with lanthanide and transition-metal ions are particularly promising due to their rich and thermally modulated electronic structures[3,18–21]. Nevertheless, previously reported systems have not utilized thermally induced enhancement of emission in any RGB channel, which could significantly improve sensitivity.

In the present work, we address this gap by demonstrating a $Ca_{19}Zn_2(PO_4)_{14}$:$Mn^{2+}$, $Ce^{3+}$ phosphor in which increasing temperature leads to a substantial thermal amplification of the emission intensity recorded in the green (G) channel of a digital camera. This effect is attributed to the synergistic action of two rare and thermally activated mechanisms: (1) a temperature-induced blueshift of the $^4T_1 \rightarrow {}^6A_1$ emission band of $Mn^{2+}$ ions, and (2) thermally assisted population of the $^4T_2$ excited state through electron trap states. The simultaneous occurrence of these mechanisms results in an intense thermal enhancement of the $Mn^{2+}$ emission intensity within the spectral range corresponding to the G channel, while the signals recorded in the B channel exhibit the expected thermal quenching. The pronounced thermal amplification of $Mn^{2+}$ luminescence is facilitated by two factors: the ability to tailor the concentration of optical traps through controlled doping, and the exceptionally long lifetime of the $^4T_1$ excited state of $Mn^{2+}$ ions, which increases the probability of thermally assisted repopulation. Notably, the thermally induced spectral shift of a transition-metal emission band is an exceptionally uncommon phenomenon among $Mn^{2+}$-doped phosphors, making this material particularly attractive for advanced thermometric applications. Additionally, the inherently low emission efficiency of $Mn^{2+}$ ions resulting from spin-forbidden electronic transitions was address through $Ce^{3+} \rightarrow Mn^{2+}$ energy transfer, which substantially enhances $Mn^{2+}$ luminescence and broadens the usable spectral window for temperature readout.

The combination of these thermally activated processes, implemented in $Ca_{19}Zn_2(PO_4)_{14}$:$Mn^{2+}$, $Ce^{3+}$, enables the development of a multimodal temperature sensor capable of both ratiometric thermal sensing and filter-free thermal imaging. The sensor operates through two distinct readout modes, offering enhanced sensitivity, high thermal contrast, and compatibility with low-cost imaging equipment. This work therefore introduces a new design principle for RGB-based luminescent thermometers that exploit thermally induced emission enhancement - an approach that has remained largely unexplored to date.

**Experimental**

*Synthesis*

The $Ca_{19(1-y)}Zn_{2(1-x)}(PO_4)_{14}$:$x$$Mn^{2+}$, $y$$Ce^{3+}$ samples were synthesized by a high-temperature solid-state reaction method. The following compounds were used as staring materials: $CaCO_3$ (99.999%), ZnO (99.99%), $NH_4H_2PO_4$ (99%), $MnCl_2 \cdot 4H_2O$ ($\geq$ 99.0%) and $Ce(NO_3)_3 \cdot 6H_2O$ (99.998%). These powders were carefully weighed and ground in an agate mortar with hexane to ensure homogeneity of the mixture. The obtained homogeneous mixtures were then put to corundum crucibles and annealed in air at 1573 K for 12 h, with heating rate of 10 K min$^{-1}$. After cooling, the obtained samples were ground in a mortar.

*Characterization*

The X-ray diffraction (XRD) analysis was performed using a PANalytical X'Pert Pro diffractometer using Ni-filtered Cu Kα radiation (V = 40 kV, I = 30 mA). Measurements were performed in the 2θ = 10 - 90º range with a 30 min measurement time.

Scanning electron microscopy (SEM) was used to verify the morphology of the samples and the distribution of its elements by Energy-Dispersive X-ray Spectroscopy (EDS). The FEI Nova NanoSEM 230 equipped with an EDAX Genesis XM4 energy dispersive spectrometer was used for measurements (V = 30 kV for SEM and V = 5 kV for EDS mapping). Samples were prepared by dispersing some amount of powder in a few drops of methanol. A drop of the resulting suspension was placed on the carbon stub and dried.

The spectroscopic analysis including emission, excitation spectra and luminescence kinetics was performed with a FLS1000 Fluorescence Spectrometer from Edinburgh Instruments, equipped with a 450 W xenon lamp and µFlash pulsed lamp as an excitation sources and an R928 photomultiplier tube from Hamamatsu as a detector. The temperature-dependent measurements were performed using a THMS 600 heating-cooling stage from Linkam (temperature stability of 0.1 K and a set point resolution of 0.1 K). Before each measurement, the temperature was stabilized for 1 min to ensure reliable readouts.

The luminescence decay profiles were fitted using double exponential function:

$$I(t) = I_0 + A_1 \cdot \exp\left(-\frac{t}{\tau_1}\right) + A_2 \cdot \exp\left(-\frac{t}{\tau_2}\right) \qquad (1)$$

where $\tau_1$ and $\tau_2$ represent decay components and $A_1$ and $A_2$ are the amplitudes of double-exponential functions. Based on the obtained results the average lifetime of the excited state was calculated as follows:

$$\tau_{avr} = \frac{A_1\tau_1^2 + A_2\tau_2^2}{A_1\tau_1 + A_2\tau_2} \qquad (2)$$

Thermoluminescence measurements, spanning the 300 - 650 K range, were conducted using a *Lexsyg Research* fully automated TL/OSL reader (Freiberg Instruments GmbH). X-ray excitation was provided by a VF-50J RTG lamp with a tungsten anode, operated under two distinct conditions: 20 kV and 0.5 mA for 5 seconds in TL glow curve acquisition, and 45 kV and 0.5 mA for 5 minutes in X-rays excited luminescence measurements. Thermoluminescence glow curves were captured using a 9235QB photomultiplier tube (ET Enterprises) over the temperature range of 300 - 650 K, employing a linear heating rate of 5 K s$^{-1}$. X-rays excited luminescence spectra were recorded by an Andor DU420A-OE CCD detector, thermoelectrically cooled to 193 K to reduce noise and enhance signal fidelity with a linear heating rate of 5 K s$^{-1}$. All experimental operations were managed via the *LexStudio 2* software. The glow curves were deconvoluted into TL components, each of them described with eq. 3 and using GlowFit software[22]:

$$I(T) = I_m \exp\left(\frac{E}{kT_m} - \frac{E}{kT}\right)\exp\left(\frac{E}{kT_m}\left(\alpha\left(\frac{E}{kT_m}\right) - \frac{T}{T_m}\exp\left(\frac{E}{kT_m} - \frac{E}{kT}\right)\alpha\left(\frac{E}{kT}\right)\right)\right) \qquad (3)$$

where *I(T)* denotes glow peak intensity, $E$ (eV) the activation energy, k (eV K$^{-1}$) the Boltzmann constant, $\alpha$ is a quotient of fourth order polynomials, and $T_m$ and $I_m$ are the temperature and the intensity of the maximum, respectively. Frequency factor, *s*, was calculated as follows:

$$\frac{\beta E}{kT_m^2} = s\exp\left(-\frac{E}{kT_m}\right) \qquad (4)$$

where $\beta$ (K s$^{-1}$) stands for the heating rate.

The digital images were taken using a Canon EOS 400D camera using a 5 s integration time, 14.3 lp/nm spatial resolution. After acquiring the luminescence images for the proof-of-concept experiment, the individual R, G and B channels were extracted and emission maps based on their intensity ratios were subsequently generated. All image processing steps were performed using IrfanView 64 (version 4.51).

**Results and discussion**

The investigated $Ca_{19}Zn_2(PO_4)_{14}$ crystallizes in a trigonal crystal system with the space group *R3cH* (161)[23–30]. In this structure, Ca, Zn, P and O atoms occupy 4, 1, 3 and 10 types of crystallographic sites, respectively. As shown in Figure 1a, $Ca_{19}Zn_2(PO_4)_{14}$ has different cation sites: three 8-fold and 6-fold coordinated Ca, 6-fold coordinated Zn , and 4-fold coordinated P site[23,26,29,30]. The small difference in ionic radii between $Ce^{3+}$ dopant ions (1.01 Å for coordination number (CN) = 6 and 1.143 Å for CN = 8) and $Ca^{2+}$ host material cation (1.00 Å for CN = 6, 1.12 Å for CN = 8) indicates that the $Ce^{3+}$ ions replace $Ca^{3+}$ sites in this structure[31]. Analogously, the $Mn^{2+}$ (0.83 Å for CN = 6) replace the $Zn^{2+}$ (0.74 Å for CN = 6) site. The comparison of the XRD patterns of $Ca_{19}Zn_2(PO_4)_{14}$ with different concentration of dopant ions with the reference data (ICSD 189197) revealed the lack of additional reflections confirming high phase purity of the obtained phosphors (Figure 1b, Figure S1, S2). However, with an increase in the dopant concentration shifting of the reflections towards smaller angles is observed which suggest the expansion of the unit cell of the $Ca_{19}Zn_2(PO_4)_{14}$ structure. This behavior is consistent with the substitution of host ions by dopant ions of larger ionic radii. The Rietveld refinement of the XRD patterns allows to trace the trend that when $Ce^{3+}$ concentration increases from 0.5 to 10%$Ce^{3+}$ the *a/b* parameters increases from 10.365 A to 10.425A (Figure 1c). Analogously, in the same concentration of dopant range the *c* parameter and volume of the unit cell (*V*) increases from 37.18 A to 37.37 A and 3460 $A^3$ to 3518 $A^3$, respectively (Figure 1d and e) (similar analysis was performed for various $Mn^{2+}$ dopant concentration Figure S3). This effect associated with the difference in ionic radii between dopant and host material cation confirms the successful incorporation of dopant ions into the $Ca_{19}Zn_2(PO_4)_{14}$ structure. The SEM analysis of the representative $Ca_{19}Zn_2(PO_4)_{14}$:2% $Mn^{2+}$,5% $Ce^{3+}$ sample revealed that the synthesized powders consist of aggregated microcrystals of around 5 μm in diameter (Figure 1f). Additionally the homogenous dopant ions distribution was confirmed by the EDX analysis as shown in Figure 1g-l.

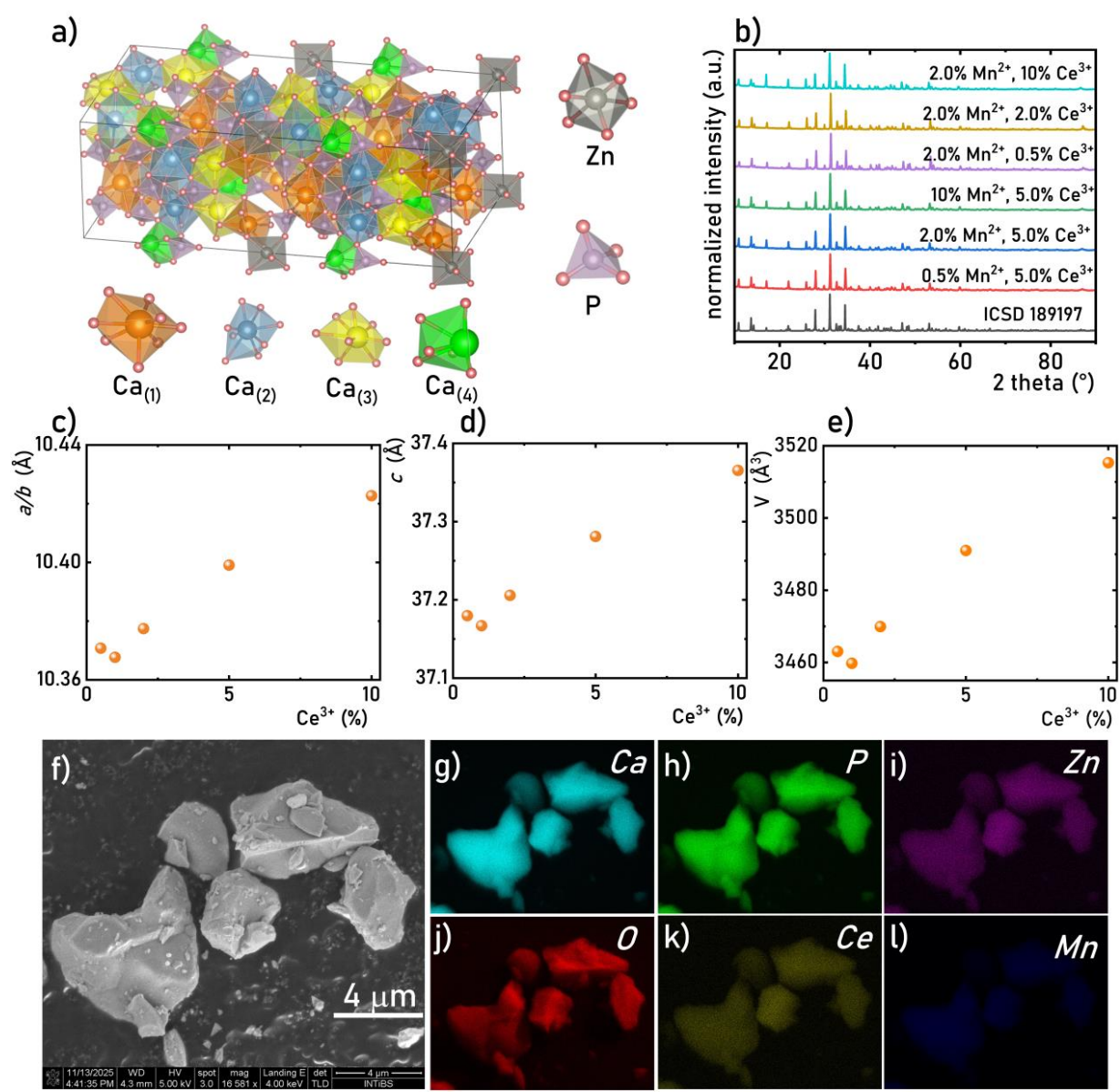

**Figure 1.** Visualization of the $Ca_{19}Zn_2(PO_4)_{14}$ structure – a); the XRD patterns of $Ca_{19}Zn_2(PO_4)_{14}$:$Mn^{2+}$, $Ce^{3+}$ with different dopant concentrations – b); the influence of the $Ce^{3+}$ ions concentration on the $a$ – c) and $c$ – d)

unit cell parameters and volume of the unit cell – e); representative SEM image of $Ca_{19}Zn_2(PO_4)_{14}$:2% $Mn^{2+}$, 5% $Ce^{3+}$ - f) and elemental distributions of Ca – g); P – h); Zn – i); O – j); Ce – k) and Mn – l).

In order to understand the spectroscopic properties of $Mn^{2+}$ and $Ce^{3+}$ ions in doped $Ca_{19}Zn_2(PO_4)_{14}$, a schematic configurational coordinate diagram is depicted in Figure 2a. The luminescent properties of $Ce^{3+}$ ions arise from electronic transitions between the *5d* excited state and the two *4f* levels $^2F_{7/2}$ and $^2F_{5/2}$. In $Ca_{19}Zn_2(PO_4)_{14}$:$Mn^{2+}$, $Ce^{3+}$, these transitions result in a broad emission band centered at approximately 365 nm (Figure 2b). The asymmetric broad band of the $Ce^{3+}$ can be decomposed into two well-fitted Gaussian profiles centering at 25831 $cm^{-1}$ and 27775 $cm^{-1}$, respectively. It is well known that the emission of $Ce^{3+}$ exhibits a doublet character due to spin-orbit splitting of the ground state ($^2F_{5/2}$ and $^2F_{7/2}$), with a theoretical energy difference of approximately 2000 $cm^{-1}$ [32,33]. In a case of doped $Ca_{19}Zn_2(PO_4)_{14}$, the energy difference is approximately 1944 $cm^{-1}$, which is close to the expected 2000 $cm^{-1}$ splitting of the 4f ground state. Under the same excitation conditions, an additional emission band is observed that corresponds to the $^4T_1 \rightarrow ^6A_1$ electronic transition of $Mn^{2+}$ ions in $Ca_{19}Zn_2(PO_4)_{14}$, revealing the $Ce^{3+} \rightarrow Mn^{2+}$ energy transfer. The appearance of the $Mn^{2+}$ emission band in the red spectral region indicates that these ions in $Ca_{19}Zn_2(PO_4)_{14}$ occupy octahedral crystallographic sites. According to the Tanabe-Sugano diagram for ions with a *3d⁵* electronic configuration, $Mn^{2+}$ ions in tetrahedral symmetry emits in the green region[34–36]; thus, red emission confirms octahedral coordination. Importantly, in our experiments the $Mn^{2+}$ luminescence in doped $Ca_{19}Zn_2(PO_4)_{14}$ cannot be obtained when the material is doped exclusively with $Mn^{2+}$ (without $Ce^{3+}$). In such samples, only a weak signal attributed to $Mn^{4+}$ is detected at 83 K (Figure S4). A comparison of the excitation spectra recorded at wavelengths corresponding to the emission of $Ce^{3+}$ ($\lambda_{em}$=360 nm) and $Mn^{2+}$ ($\lambda_{em}$=665 nm) shows a clear difference in the absorption cross sections of these ions (Figure 2c, see also Figure S5-S8). For $Ce^{3+}$ emission, distinct excitation bands are observed between 260 and 330 nm, which correspond to transitions from the $^2F_{5/2}$ ground state to the *5d* levels. For $Mn^{2+}$ emission, the same excitation bands appear, together with an additional band at 365 nm associated with the $^6A_1 \rightarrow {}^4T_1$ electronic transition. The presence of $Ce^{3+}$ excitation bands in the $Mn^{2+}$ excitation spectrum directly confirms the occurrence of $Ce^{3+} \rightarrow Mn^{2+}$ energy transfer. Moreover, the $Ce^{3+}$-related excitation bands exhibit substantially higher intensity than those associated with $Mn^{2+}$ ions, demonstrating that $Ce^{3+} \rightarrow Mn^{2+}$ energy transfer is a far more efficient mechanism for populating the $^4T_1$ excited state of $Mn^{2+}$ ions compared to direct excitation. This efficiency is largely due to the fact that the *4f* $\rightarrow$ *5d* transition in $Ce^{3+}$ ions is an allowed electronic transition. The relative intensity of the emission bands of $Ce^{3+}$ and $Mn^{2+}$ can be tuned by adjusting the concentrations of the dopant ions (Figure 2d, Figure S9, S10). This effect is clearly illustrated in samples containing a constant 5% $Ce^{3+}$ concentration and varying $Mn^{2+}$ concentration. For a low 0.25% $Mn^{2+}$, the emission spectrum of $Ca_{19}Zn_2(PO_4)_{14}:Mn^{2+}, Ce^{3+}$ is dominated by $Ce^{3+}$ emission band. As the $Mn^{2+}$ concentration increases, the $Mn^{2+}$ emission band becomes progressively more intense, which can be attributed to the increased number of $Mn^{2+}$ emission centers. However, a similar effect is observed when the $Mn^{2+}$ concentration is held constant and the $Ce^{3+}$ concentration is increased, demonstrating that the reduction in $Ce^{3+}$ emission relative to $Mn^{2+}$ results from more efficient energy transfer at higher dopant concentrations. Increasing the amount of either ion decreases the average distance between interacting ions, thereby facilitating interionic energy transfer. Consequently, the normalized emission intensity ratio of $Ce^{3+}$ to $Mn^{2+}$ decreases from approximately 14 for 0.25% $Mn^{2+}$,5% $Ce^{3+}$ to about 1.2 for 10% $Mn^{2+}$,5% $Ce^{3+}$ (Figure 2e). Because the emission bands of the two ions are well separated spectrally and lie in different spectral regions, adjusting the dopant composition produces noticeable changes in the color of the emitted light. An analysis performed at 83 K using CIE1931 chromaticity coordinates shows that the emission color shifts from violet for 0.25% $Mn^{2+}$,5% $Ce^{3+}$ to reddish-pink for 10% $Mn^{2+}$,5% $Ce^{3+}$ (Figure 2f). This tunability is highly advantageous for practical applications, as it allows the design of phosphors with precisely tailored emission colors through appropriate selection of dopant concentrations.

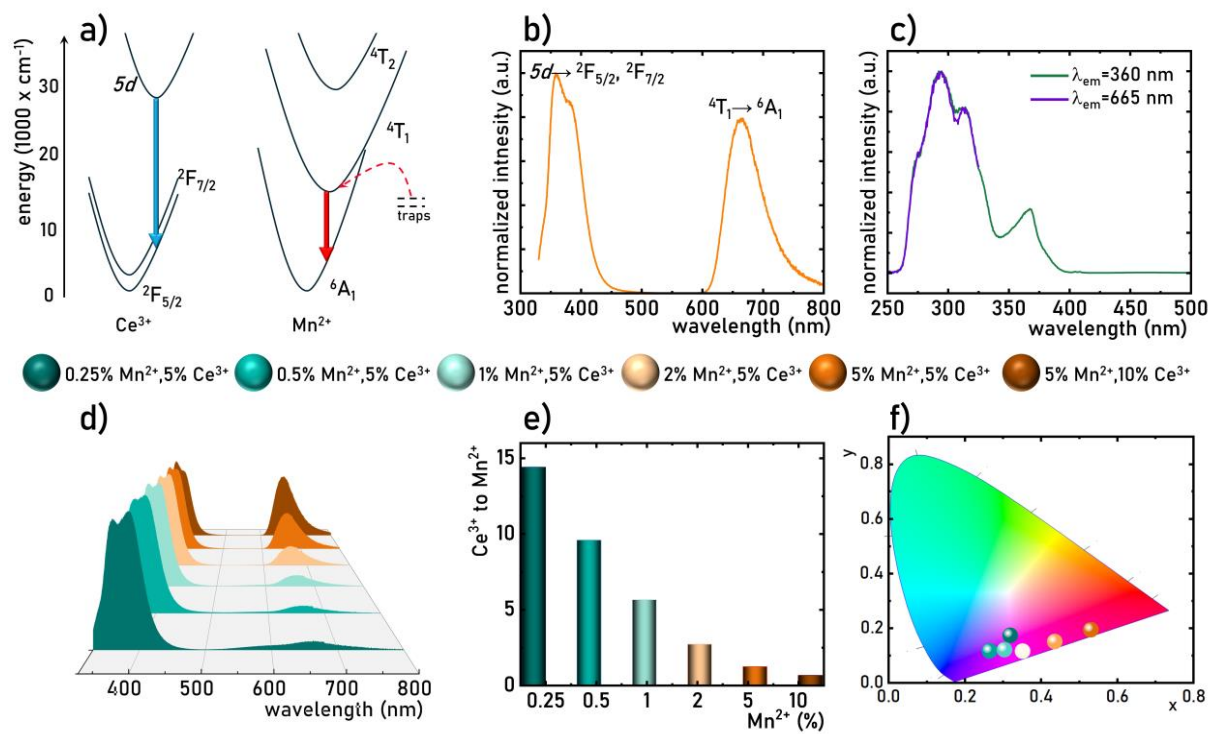

**Figure 2.** Simplified configurational coordination diagram of $Ce^{3+}$ and $Mn^{2+}$ ions – a); representative emission spectrum of $Ca_{19}Zn_2(PO_4)_{14}$:5% $Mn^{2+}$, 10% $Ce^{3+}$ measured at 83 K – b); comparison of excitation spectra of $Ca_{19}Zn_2(PO_4)_{14}$:10% $Mn^{2+}$, 5% $Ce^{3+}$ measured at 83 K at two different emission wavelengths – c); emission spectra of $Ca_{19}Zn_2(PO_4)_{14}$:5% $Mn^{2+}$, $Ce^{3+}$ with different $Ce^{3+}$ ions concentration measured at 83 K– d); the influence of $Ce^{3+}$ ions concentration on the $Ce^{3+}$ to $Mn^{2+}$ emission intensity ratio at 83 K – e); and corresponding CIE1931 chromatic coordinates – f).

The measurements of the $Ca_{19}Zn_2(PO_4)_{14}$:$Mn^{2+}$, $Ce^{3+}$ emission spectra as a function of temperature revealed several characteristic effects (Figure 3a, Figure S11-21). As the temperature increased, the intensity of the $5d \rightarrow 4f$ transition of the $Ce^{3+}$ band gradually decreased. In contrast, the $^4T_1 \rightarrow \,^6A_1$ $Mn^{2+}$ band exhibited a spectral blueshift accompanied by an increase in intensity with rising temperature. A thermal blueshift of the $Mn^{2+}$ band has previously been reported for other phosphors and is associated with thermal population of higher vibronic levels of the $^4T_1$ state[34,36–38]. A comparative analysis of materials with different $Ce^{3+}$ and $Mn^{2+}$ concentrations showed that the rate of the spectral shift is independent of the dopant concentration (Figure 3b-e), with the barycenter of the $Mn^{2+}$ band shifting from approximately 668 nm at 83 K to 638 nm at 703 K, supporting this interpretation (Figure 3f). Although this type of behavior is often attributed to the presence of several nonequivalent crystallographic sites occupied by the dopant ions, measurements of luminescence decay kinetics (Figure S22) and excitation spectra at different detection wavelengths (Figure S23) indicated no such differences in $Ca_{19}Zn_2(PO_4)_{14}:Mn^{2+}, Ce^{3+}$. These findings suggest the presence of a single $Mn^{2+}$ emission center in this host material. However, the thermal enhancement of $Mn^{2+}$ emission intensity is an unusual effect. Importantly, at low $Mn^{2+}$ concentrations this enhancement is absent; only the spectral shift of the band is observed (Figure 3b-e). Increasing the $Mn^{2+}$ concentration leads to a progressive enhancement of the thermally induced increase in $Mn^{2+}$ intensity. Moreover, increasing the $Ce^{3+}$ concentration while keeping the $Mn^{2+}$ concentration constant results in a similar effect, indicating that this behavior may be linked to $Ce^{3+} \rightarrow Mn^{2+}$ energy transfer (Figure S16-21). In this scenario, increasing the concentration of each ion effectively reduces the average interionic distance, thereby facilitating energy transfer between them. A comparison of the thermal quenching behavior of a phosphor doped only with $Ce^{3+}$, which exhibits higher thermal stability than the $Ce^{3+}$, $Mn^{2+}$ co-doped material, also supports this hypothesis (Figure S24). However, a similar increase in $Mn^{2+}$ emission intensity was observed upon direct excitation of $Mn^{2+}$ in the $^4T_1 \rightarrow ^6A_1$ band ($\lambda_{exc}$=365 nm), which rules out energy transfer as the underlying mechanism (Figure S25). Analysis based on the Tanabe-Sugano diagram for ions with a *$3d^5$* electronic configuration in octahedral symmetry indicates that the simultaneous increase in $Mn^{2+}$ emission intensity and the spectral blueshift could be related to a thermally induced decrease in the crystal field strength affecting $Mn^{2+}$ ions[39]. Such an effect, however, would require strong thermal expansion of the unit cell or at least of the $Mn^{2+}$-occupied octahedra. However, this was not observed for $Ca_{19}Zn_2(PO_4)_{14}$ [23,30,40]. Furthermore, an increase in the concentrations of $Mn^{2+}$ and $Ce^{3+}$ would not be expected to enhance this effect. Therefore, the thermal increase in $Mn^{2+}$ luminescence intensity is most likely linked to the presence of electron traps located below the excited state of $Mn^{2+}$ created by the charge mismatch between $Ce^{3+}$ ions and the $Ca^{2+}$ sites they occupy. The resulting hole traps generate defect states that, when located near $Mn^{2+}$ ions, can serve as an additional channel for populating the $Mn^{2+}$ excited state. Increasing the temperature promotes thermal release of trapped carriers, thereby enhancing the population of the $Mn^{2+}$ excited state. Although this mechanism is typically associated with persistent luminescence and manifests primarily in modified decay kinetics[41–43] rather than intensity enhancement, the long lifetime of the $^4T_1$ state of $Mn^{2+}$ favors a thermally induced increase in its emission intensity. Analysis of the thermal evolution of $Ce^{3+}$ luminescence revealed reduced thermal stability of the $Ce^{3+}$ emission with increasing dopant concentration, consistent with enhanced $Ce^{3+}\rightarrow Mn^{2+}$ energy transfer (Figure 3g). On the other hand, the thermally induced enhancement of $Mn^{2+}$ emission results in an intensity increase to approximately 120% of the value at 83 K for a 0.5% $Ce^{3+}$ ions. A higher $Ce^{3+}$ concentration leads to an increased rate of thermal enhancement, culminating in a 280% increase in $Mn^{2+}$ intensity at 583 K for a 10% $Ce^{3+}$, followed by a slight decrease at higher temperatures (the similar changes are observed when $Mn^{2+}$ ions concentration increases - see Figure S26, S27). This opposite thermal trend of $Ce^{3+}$ and $Mn^{2+}$ emissions enables ratiometric temperature sensing based on the following definition:

$$LIR_1 = \frac{Ce^{3+}}{Mn^{2+}} = \frac{\int_{350nm}^{500nm} (5d \rightarrow {}^4F_{5/2}, F_{7/2}) d\lambda}{\int_{550nm}^{800nm} ({}^4T_1 \rightarrow {}^6A_1) d\lambda} \quad (5)$$

For the composition $Ca_{19}Zn_2(PO_4)_{14}$:2% $Mn^{2+}$, 0.5% $Ce^{3+}$, the $LIR_1$ value decreases by approximately 40% over the entire temperature range examined, due to only minor changes in both $Ce^{3+}$ and $Mn^{2+}$ emissions (Figure 3i, Figure S28). However, the magnitude of $LIR_1$ variation increases with increasing dopant concentrations. Consequently, for the 2% $Mn^{2+}$ and 10% $Ce^{3+}$, $LIR_1$ decreases by 80% at 600 K. To quantify these temperature-induced changes in $LIR_1$, the relative thermal sensitivity of the luminescent thermometer was defined as follows:

$$S_R = \frac{1}{LIR}\frac{\Delta LIR}{\Delta T} \cdot 100\% \qquad (6)$$

where $\Delta LIR$ stands for the change of $LIR$ corresponding to the change in temperature by $\Delta T$. The difference in the thermal behavior of $LIR_1$ for $Ca_{19}Zn_2(PO_4)_{14}$:2% $Mn^{2+}$, $Ce^{3+}$ and varying $Ce^{3+}$ concentrations is reflected in the relative sensitivity, which increases with $Ce^{3+}$ concentration and reaches a maximum of $S_R=0.42\%K^{-1}$ at 500 K for the material with $Ca_{19}Zn_2(PO_4)_{14}$:2% $Mn^{2+}$, 10% $Ce^{3+}$ (Figure 3j).

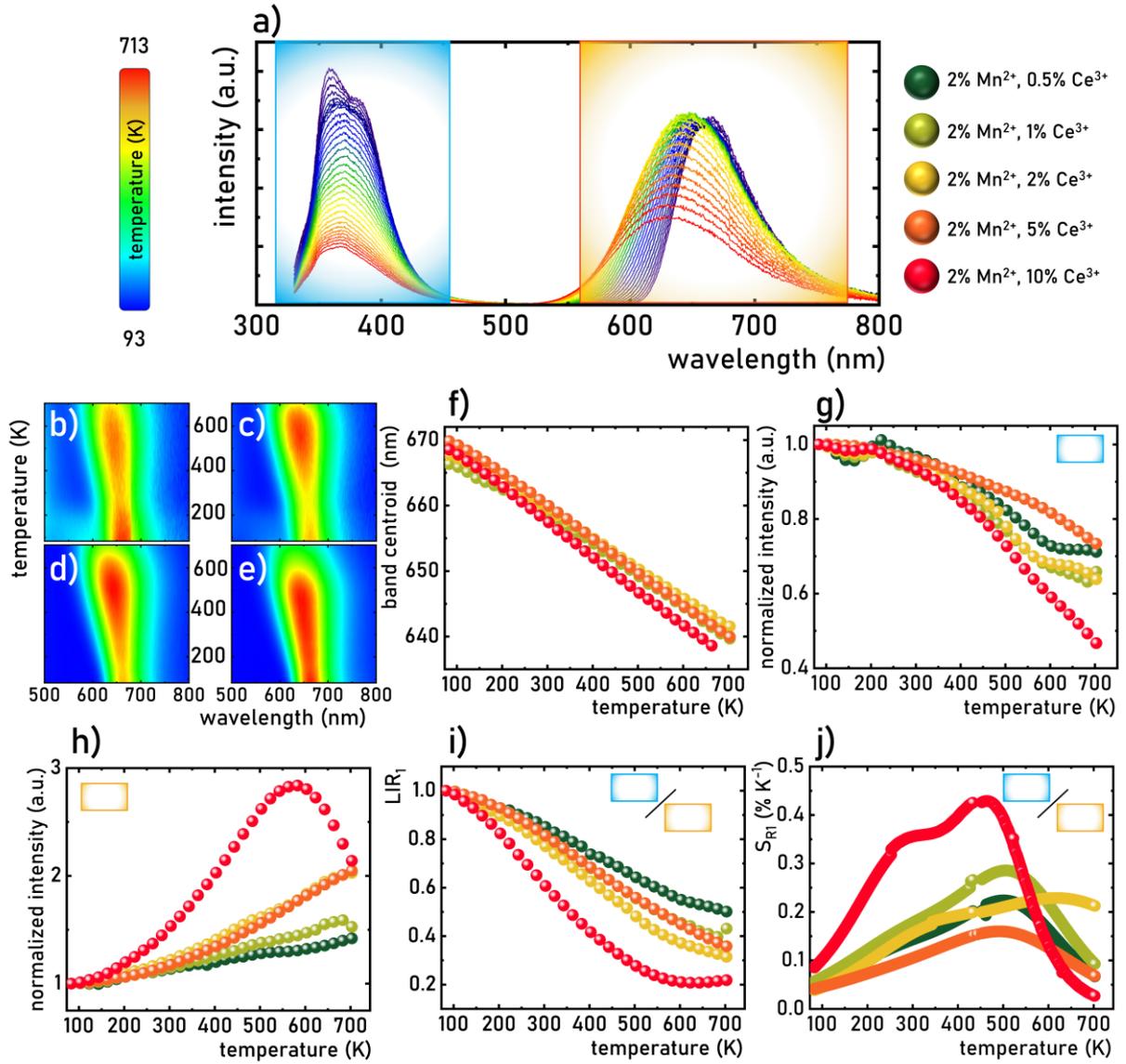

**Figure 3.** Emission spectra of $Ca_{19}Zn_2(PO_4)_{14}$:5% $Mn^{2+}$, 10% $Ce^{3+}$ measured as a function of temperature – a); thermal map of the emission band of $Mn^{2+}$ for $Ca_{19}Zn_2(PO_4)_{14}$:2% $Mn^{2+}$, $Ce^{3+}$ with 0.5% $Ce^{3+}$ – b); 1% $Ce^{3+}$ – c); 5% $Ce^{3+}$ – d); 10% $Ce^{3+}$ – e); thermal dependence of emission band barycenter of $Mn^{2+}$ ions – f); thermal dependence of integrated emission intensity of $Ce^{3+}$ ions for different concentration of $Ce^{3+}$ ions – g); thermal dependence of integrated emission intensity of $Mn^{2+}$ ions for different concentration of $Ce^{3+}$ ions – h); thermal dependence of $LIR_1$ – i); and corresponding $S_{R1}$ – j).

In order to verify the presence of optical traps in the analyzed phosphor the thermoluminescence (TL) measurements following X-ray irradiation were performed for the series of materials $Ca_{19}Zn_2(PO_4)_{14}$:$Mn^{2+}$, $Ce^{3+}$. As shown in the thermoluminescence glow curves, normalized to the mass of 1 g of the material (Figure 4a) the TL response after X-ray exposure is essentially negligible for the sample activated with $Mn^{2+}$ ions. In contrast, the remaining samples exhibit qualitatively similar glow-curve profiles in the 300-600 K temperature range, with a pronounced intensity enhancement between 300-420 K. This result is the direct confirmation about the presence of optical traps in samples doped or co-doped with $Ce^{3+}$ ions. The obtained curves were deconvoluted into individual components using the GlowFit software package. In the analyzed systems achieving high-quality fits (FOM = 3–5%) required the application of ~9 discrete components with distinct activation energies (*E*). The resulting trap parameters are summarized in **Table S1**. The trap depths fall within the 0.7–1.7 eV range and are comparable for all investigated materials. Notably, the derived frequency factors (s) predominantly lie within $10^{11}$-$10^{14}$ s$^{-1}$, consistent with first-order TL kinetics[44]. For each sample, substantial overlap between the deconvoluted peaks is evident, an effect widely recognized in literature as one of the most challenging scenarios for kinetic analysis. Thermal evolution of thermoluminescence glow curve of $Ca_{19}Zn_2(PO_4)_{14}$:2% $Mn^{2+}$, 10% $Ce^{3+}$ revealed that across the entire temperature interval in which TL is observed, the spectrum consists of $Mn^{2+}$ and $Ce^{3+}$ emission bands (Figure 4c). Taken together, these results highlight the complex trap structure and multi-center recombination dynamics in $Ca_{19}Zn_2(PO_4)_{14}$:2% $Mn^{2+}$, 10% $Ce^{3+}$ -based phosphors, underscoring their potential utility as model systems for studying interplay between activator ions and competing trapping pathways in luminescent materials.

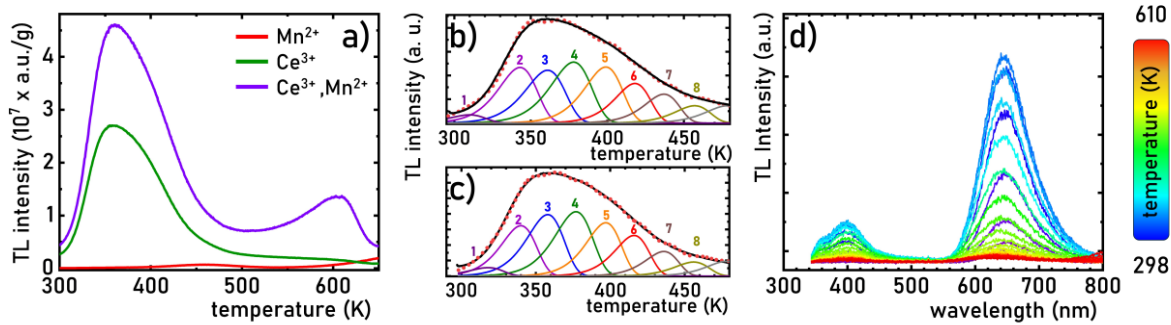

**Figure 4**. Thermoluminescence glow curves measured for $Ca_{19}Zn_2(PO_4)_{14}$:10%$Mn^{2+}$, $Ca_{19}Zn_2(PO_4)_{14}$: 10% $Ce^{3+}$ and $Ca_{19}Zn_2(PO_4)_{14}$:2% $Mn^{2+}$, 10% $Ce^{3+}$ -a ); the deconvolution of the thermoluminescence curves for $Ca_{19}Zn_2(PO_4)_{14}$: 10% $Ce^{3+}$ - b) and $Ca_{19}Zn_2(PO_4)_{14}$:2% $Mn^{2+}$, 10% $Ce^{3+}$ - c); thermoluminescence glow curves of $Ca_{19}Zn_2(PO_4)_{14}$: 10% $Ce^{3+}$ measured as a function of temperature – d).

The thermally-induced spectral shift of the emission band of $Mn^{2+}$ ions in $Ca_{19}Zn_2(PO_4)_{14}$:$Mn^{2+}$, $Ce^{3+}$ represents an uncommon behavior, rarely reported for phosphors doped with transition metal ions. Typically, increasing temperature results in a decrease in the emission band intensity without a significant change in its spectral position[3,21,45–47]. The unusual thermal behavior of the $^4T_1 \rightarrow {}^6A_1$ $Mn^{2+}$ band enables the implementation of a ratiometric temperature readout based on the ratio of luminescence intensities recorded within two spectral ranges[48–50], as indicated in Figure 5a (*G1*: 560-600 nm, *G2*: 660-750 nm). Within these spectral windows, the blueshift of the $Mn^{2+}$ emission band causes a decrease in the luminescence intensity recorded in *G2*, while simultaneously increasing the intensity in *G1* (see also Figure S29). The selected spectral ranges were chosen empirically to optimize temperature-dependent response. As previously demonstrated, the rate of the $Mn^{2+}$ spectral shift is independent of dopant ion concentration (Figure 3). However, the simultaneous occurrence of a spectral shift and thermal enhancement of luminescence intensity should lead to a more pronounced thermal increase in the signal recorded in *G1* with higher dopant concentrations. Comparative analysis of the thermal evolution of the *G1* signal (Figure 5b) confirms this assumption. For the sample containing 2% $Mn^{2+}$ and 10% $Ce^{3+}$, the signal intensity increased nearly 100-fold over the investigated temperature range. In contrast, at low $Ce^{3+}$ concentrations, increasing temperature leads to a reduction in the luminescence intensity recorded in *G2* (Figure 5c). With increasing $Ce^{3+}$ concentration, the rate of thermal quenching in *G2* becomes less pronounced, and for concentrations above 2% $Ce^{3+}$, a slight thermal increase in *G2* intensity is observed. The maximum increase in *G2*, approximately 1.5-fold, was recorded for 5% $Ce^{3+}$. Further increases in $Ce^{3+}$ concentration did not result in significant additional enhancement. The thermal increase in *G2* intensity primarily originates from the overall thermal enhancement of the $Mn^{2+}$ emission. Nevertheless, compared to the pronounced changes observed in *G1*, the contribution of this effect to the *G2* signal remains minor. Consequently, the ratio of the luminescence intensities recorded in *G1* and *G2* can be defined as a ratiometric temperature parameter, $LIR_2$:

$$LIR_2 = \frac{Mn^{2+}(G1)}{Mn^{2+}(G2)} = \frac{\int_{560nm}^{600nm}\left(^4T_1 \to {}^6A_1\right)d\lambda}{\int_{660nm}^{720nm}\left(^4T_1 \to {}^6A_1\right)d\lambda} \qquad (7)$$

As shown in Figure 5d, increasing temperature leads to a consistent rise in $LIR_2$ across all dopant concentrations. The most pronounced, nearly 110-fold increase, was observed for the sample containing 10% $Ce^{3+}$. This behavior is reflected in the $S_{R2}$, which reaches a maximum value of 1.8% $K^{-1}$ at 83 K for 10% $Ce^{3+}$ (Figure 5e). However, no clear correlation was found between dopant concentration and the maximum $S_{R2}$ value. In ratiometric luminescent thermometry, the inclusion of signals exhibiting maximum and opposite thermal monotonicity enhances $S_R$. As noted above, the greatest reduction in *G2* intensity was observed for the sample containing 2% $Mn^{2+}$ and 0.5% $Ce^{3+}$, where the signal decreased to approximately 20% of its initial value. In contrast, previous studies (Figure 3) have shown that under the same thermal conditions, the luminescence intensity of $Ce^{3+}$ ions decreases by nearly 60%. Therefore, the luminescence intensity ratio $LIR_3$ was defined to combine these effects:

$$LIR_3 = \frac{Mn^{2+}(G1)}{Ce^{3+}} = \frac{\int_{560nm}^{600nm}\left(^4T_1 \to {^6A_1}\right)d\lambda}{\int_{350nm}^{500nm}\left(5d \to {^4F_{5/2}, F_{7/2}}\right)d\lambda} \quad (8)$$

The highest dynamics of $LIR_3$ variation were again observed for the sample with 10% $Ce^{3+}$, where $LIR_3$ increased more than 200-fold across the investigated temperature range (Figure 5f, Figure S29). Consequently, $S_{R3}$ achieved a significantly higher value of 2.7% $K^{-1}$ compared to $S_{R2}$ (Figure 5g). It is important to emphasize that all spectral regions analyzed here are well separated spectrally, which facilitates the use of appropriate bandpass filters to isolate them, thereby simplifying practical temperature readout.

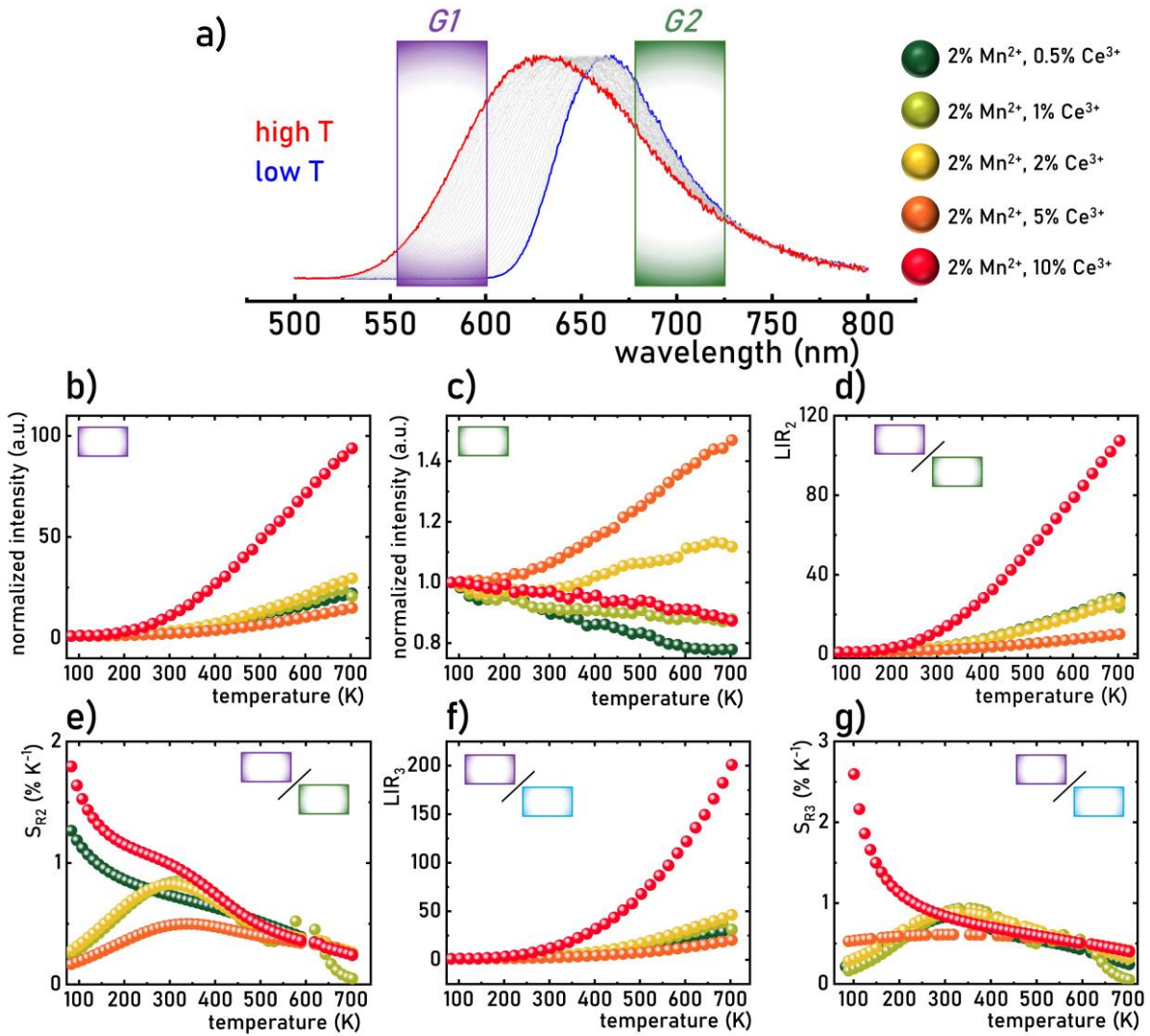

**Figure 5.** Representative thermal evolution of normalized emission band of $Mn^{2+}$ ions with the spectral ranges used for the calculations – a); thermal evolution of integrated emission intensity calculated in the 560-600 nm spectral range – b) and in the in the 660-720 nm spectral range – c);thermal evolution of $LIR_2$ – d); and $LIR_3$ – e) and corresponding $S_{R2}$ – f) and $S_{R3}$ – g).

The thermal variations described above in the luminescence intensity ratio of $Ce^{3+}$ and $Mn^{2+}$ ions in $Ca_{19}Zn_2(PO_4)_{14}$:$Mn^{2+}$, $Ce^{3+}$ lead to a temperature-dependent modulation of the emitted light color. Representative images recorded for $Ca_{19}Zn_2(PO_4)_{14}$:2%$Mn^{2+}$, 10%$Ce^{3+}$ at different temperatures clearly show that the emission color shifts from violet at 93 K to an intense pink at 493 K (Figure 6a). These changes can also be exploited for visual temperature readout. Calculations of the CIE1931 chromaticity coordinates for $Ca_{19}Zn_2(PO_4)_{14}$:$Mn^{2+}$, $Ce^{3+}$ with different $Ce^{3+}$ concentrations confirm that these color changes are monotonic across the entire analyzed temperature range (Figure 6b). The broadest color variation is observed for samples with high $Ce^{3+}$ content, which results from the previously discussed thermal enhancement of $Mn^{2+}$ emission in these compositions. A detailed analysis of the thermal evolution of the $x$ and $y$ chromaticity coordinates further supports this observation (Figure 6c). For samples with low $Ce^{3+}$ concentrations, the $x$ coordinate measured at 93 K is relatively high and changes by approximately 0.1 across the analyzed temperature range. As the $Ce^{3+}$ concentration increases, the initial $x$ value at 93 K decreases, while its thermal variability is enhanced, reaching up to 0.2 of change for a 10% $Ce^{3+}$. On the other hand, the $y$ coordinate exhibits smaller thermal changes overall, although the same general trend is observed. Consequently, both $S_{Rx}$ and $S_{Ry}$ increase with rising $Ce^{3+}$ concentration, reaching maximum values of $S_{Rx} = 0.13\%$ $K^{-1}$ and $S_{Ry} = 0.22\%$ $K^{-1}$ at 500 K for a composition containing 10% $Ce^{3+}$ (Figure 6d and e). The maximum $S_{Rx}$ and $S_{Ry}$ values vary monotonically with dopant concentration (Figure 6f and g). However, while $S_{Rx}$ increases continuously across the entire concentration range, the most significant changes in $S_{Ry}$ occur up to 2% $Ce^{3+}$, above which a saturation effect is observed and further increases in $Ce^{3+}$ concentration do not produce substantial changes in $S_{Rymax}$.

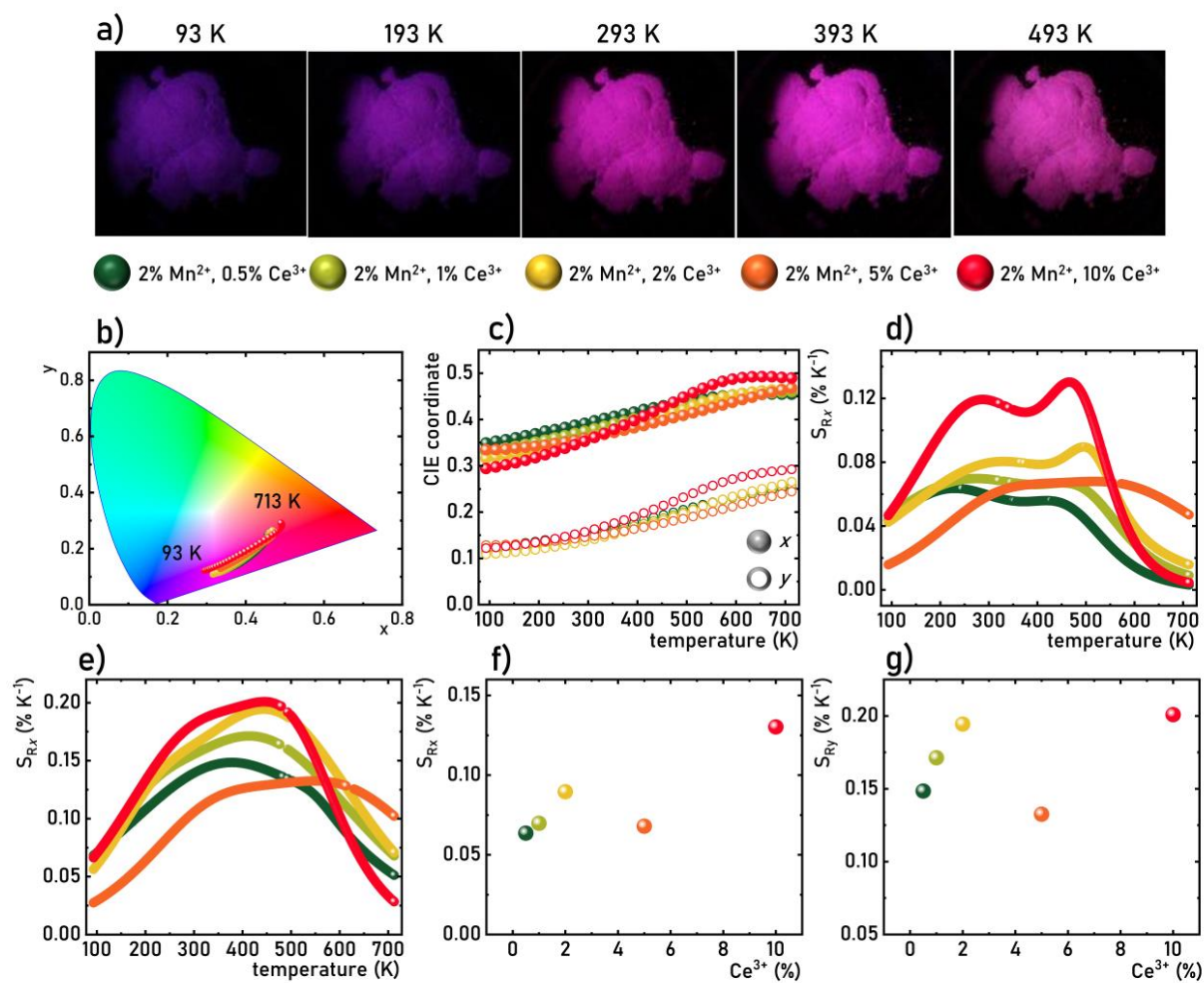

**Figure 6**. Representative photos of the luminescence of $Ca_{19}Zn_2(PO_4)_{14}$:2%$Mn^{2+}$, 10%$Ce^{3+}$ obtained at different temperatures - a); the thermal evolution of chromatic CIE1931 coordinates for of $Ca_{19}Zn_2(PO_4)_{14}$:$Mn^{2+}$, $Ce^{3+}$ for different dopant concentrations - b); the influence of temperature on $x$ and $y$ chromatic coordinates - c); and corresponding $S_{Rx}$ - d); and $S_{Ry}$ - e); the influence of dopant concentration on the $S_{Rxmax}$ - f); and $S_{Rymax}$ - g).

Although the thermally induced color variation of the emitted light in $Ca_{19}Zn_2(PO_4)_{14}$:$Mn^{2+}$, $Ce^{3+}$ is not visually pronounced, analysis of the emission spectra recorded at different temperatures in relation to the spectral sensitivity profiles of a digital camera in the blue (B), green (G), and red (R) channels reveals a significant and exploitable phenomenon. The thermally induced blueshift of the $Mn^{2+}$ emission band, accompanied by a simultaneous increase in its intensity with rising temperature, results in a marked enhancement of the signal recorded in the G channel (Figure 7a). In contrast, the thermal quenching of the $Ce^{3+}$ emission band leads to a progressive reduction of the intensity detected in the B channel. A minor increase in the R-channel signal at elevated temperatures originates from the aforementioned thermal enhancement of the $Mn^{2+}$ emission. The mutually opposite and monotonic thermal behavior of the intensities recorded in the G and B channels provides an effective ratiometric parameter for temperature determination. This is particularly advantageous from an application-oriented perspective, as the method requires acquisition of only a single image and subsequent computation of the intensity ratio between the two channels. This eliminates the need for optical filters, substantially reduces the cost and complexity of the measurement system, and facilitates straightforward and fully automatable data processing. To validate the performance of this filter-free thermal imaging approach, a circular metal plate of thickness 2 mm was coated with 2%$Mn^{2+}$, 10%$Ce^{3+}$ phosphor and positioned above an electric heater (Figure 7b). A digital camera and an infrared camera were placed above the sample to enable parallel acquisition of luminescence and reference thermal images. The phosphor layer was excited using 315 nm wavelength, and luminescence images were recorded at 5 s intervals after activating the heating element (Figure S30). Intensity maps were extracted for each of the camera channels (Figure S31-S33), and G/B ratio maps were subsequently constructed (Figure S34, see also Figure S35 and S36). Using a previously established calibration function describing the temperature dependence of the G/B ratio (Figure 7c), the maps were converted into thermal distributions (Figure 7d). While the raw luminescence images exhibit only subtle chromatic changes that are difficult to discern visually, the G/B ratio maps provide clear and quantitative insight into the spatial and temporal evolution of the temperature field on the plate. The resulting thermal maps show excellent agreement with those acquired using the infrared camera (Figure S37), enabling precise monitoring of thermal gradients, cross-sectional temperature profiles, and the temporal development of both the maximum and mean plate temperature (Figure 7e). Comparative analysis of B/R, G/R, and G/B ratios confirms the clear superiority of the G/B metric, which arises from the largest thermally induced changes being concentrated in the G channel (Figure S31-33). Crucially, this high thermometric performance originates from the synergy of two thermally activated effects: the spectral shift of the $Mn^{2+}$ emission band and the enhancement of its intensity resulting from thermally activated population of the excited state via trap levels. Although demonstrated here for the first time in $Ca_{19}Zn_2(PO_4)_{14}:Mn^{2+}, Ce^{3+}$, this combined mechanism is likely to be applicable to a broader class of $Mn^{2+}$-containing phosphors, suggesting a generalizable strategy for developing efficient materials for filter-free luminescent thermal imaging.

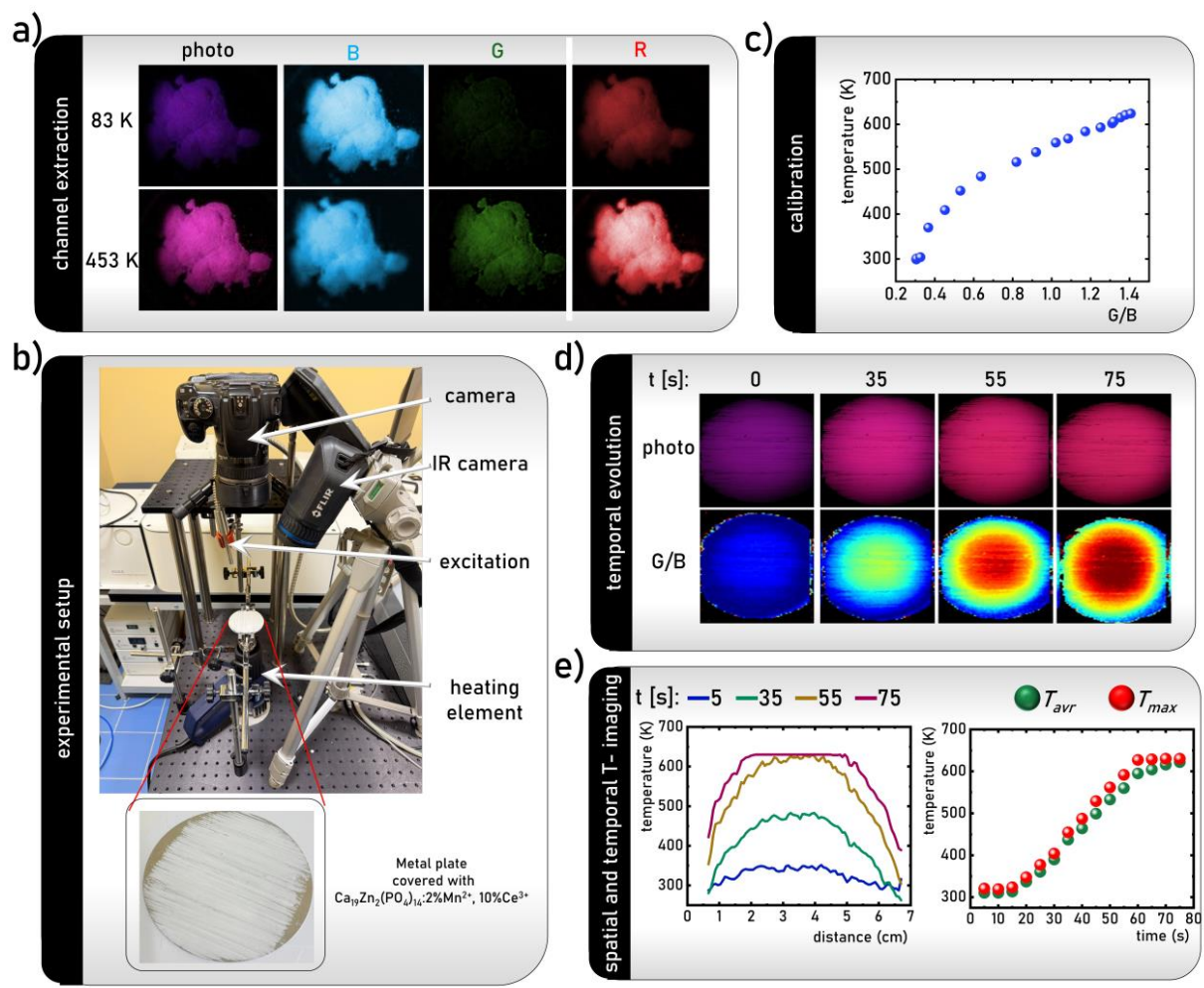

**Figure 7**. The photos of the luminescence of $Ca_{19}Zn_2(PO_4)_{14}$:2%$Mn^{2+}$, 10%$Ce^{3+}$ captured at two different temperatures with the extracted intensities maps recorded in blue (B), green (G) and red (R) channels – a); experimental setup used for thermal imaging experiment with $Ca_{19}Zn_2(PO_4)_{14}$:2%$Mn^{2+}$, 10%$Ce^{3+}$ - b); calibration curve: the dependence of the temperature on the G/B ratio - c); the images of the luminescence of $Ca_{19}Zn_2(PO_4)_{14}$:2%$Mn^{2+}$, 10%$Ce^{3+}$ captured as a function of time after heat exposure - d); the thermal cross section of the plate covered with phosphors at different times and temporal dependence of an average ($T_{avr}$) and maximal ($T_{max}$) temperature of the plate determined using luminescence thermometer- e).

**Conclusions**

In this work, a comprehensive investigation of the spectroscopic properties of $Ca_{19}Zn_2(PO_4)_{14}$:$Mn^{2+}$, $Ce^{3+}$ was carried out as a function of dopant concentration and temperature to evaluate its suitability for luminescent thermometry in both sensing and thermal imaging applications. The material exhibits characteristic emission bands of $Ce^{3+}$ ions and $Mn^{2+}$ ions, with $5d{\rightarrow}4f$ $Ce^{3+}$ luminescence centered at approximately 360 nm and $^4T_1{\rightarrow}^6A_1$ $Mn^{2+}$ emission near 660 nm. It was demonstrated that the $Ce^{3+}{\rightarrow}Mn^{2+}$ energy transfer constitutes the dominant mechanism for populating the $Mn^{2+}$ excited state, and is significantly more efficient than direct $d{\rightarrow}d$ excitation into the electronic transition associated with the $^6A_1{\rightarrow}^4T_1$ electronic transition. Increasing the concentrations of $Ce^{3+}$ and $Mn^{2+}$ ions enhances the efficiency of this energy transfer process. As a result, the relative intensity of the $Mn^{2+}$ emission band increases in respect to the $Ce^{3+}$ emission, producing a progressive shift in the emitted light color from violet at low dopant concentrations to pinkish-red at high concentrations. With increasing temperature, the $Ce^{3+}$ emission band undergoes gradual thermal quenching. In contrast, the $Mn^{2+}$ emission exhibits two distinct thermally driven effects. Regardless of dopant concentration, the $Mn^{2+}$ emission band undergoes a thermally induced spectral blueshift attributed to thermalization of the higher-lying vibronic components of the $^4T_1$ excited state. Additionally, for high dopant concentrations, an increase in $Mn^{2+}$ emission intensity occurs due to thermally activated population of the $^4T_1$ level via depopulation of optical trap states. The presence of charge mismatch between the $Ca^{2+}$ host lattice and the $Ce^{3+}$ dopant ions leads to the formation of electron trap levels. Therefore an increase in the dopant concentration reduces the average distance between $Mn^{2+}$ ions and these trap sites, thereby intensifying the thermally activated enhancement of $Mn^{2+}$ luminescence.

The distinct thermal behaviors of the cerium and manganese emissions enabled the development of several luminescent thermometric modes. A ratiometric thermometer based on the $Ce^{3+}$ to $Mn^{2+}$ emission intensity ratio achieved a maximum relative sensitivity of 0.42% $K^{-1}$ at 500 K for a composition containing 2% $Mn^{2+}$ and 10% $Ce^{3+}$. Furthermore, thermal modifications of the $Mn^{2+}$ emission band shape allowed the development of a second mode of ratiometric thermometer using the intensity ratio of two spectral regions of the manganese band, yielding a maximum sensitivity of 1.8% K$^{-1}$ at 93 K. However, combining the thermally induced blueshift of the Mn$^{2+}$ emission with the strong thermal quenching of Ce$^{3+}$ luminescence enabled a further enhancement in performance, resulting in a maximum relative sensitivity of 2.7% K$^{-1}$ for the composition containing 2% Mn$^{2+}$ and 10% Ce$^{3+}$.

While these readout modes confirm the strong potential of Ca$_{19}$Zn$_2$(PO$_4$)$_{14}$:Mn$^{2+}$, Ce$^{3+}$ for luminescent thermal sensing, the thermally induced spectral shift of the Mn$^{2+}$ emission band plays an key role in thermal imaging. It produces a pronounced increase in the signal recorded in the green channel of a digital camera, accompanied by a decrease in the blue and red channels. This behavior was exploited to demonstrate filter-free thermal imaging, based on the ratio of the camera's RGB channel intensities. The results show that Ca$_{19}$Zn$_2$(PO$_4$)$_{14}$:Mn$^{2+}$, Ce$^{3+}$ enables rapid and remote visualization of dynamic thermal changes using a standard digital camera without the need for optical filters or specialized instrumentation. Overall, the synergy between the intrinsic spectroscopic properties of manganese ions and the thermally activated population of their excited states from optical trap levels positions Ca$_{19}$Zn$_2$(PO$_4$)$_{14}$:Mn$^{2+}$, Ce$^{3+}$ as a highly promising material for next-generation luminescent thermometry and low-cost thermal imaging technologies.

**Acknowledgements**

This work was supported by the Foundation for Polish Science under First Team FENG.02.02-IP.05-0018/23 project with funds from the 2nd Priority of the Program European Funds for Modern Economy 2021-2027 (FENG). Authors would like to acknodledge dr Damian Szymanski for help in the EDX analysis.

**References**

[1]  X.D. Wang, O.S. Wolfbeis, R.J. Meier, Luminescent probes and sensors for temperature, Chem Soc Rev 42 (2013) 7834–7869. https://doi.org/10.1039/C3CS60102A.

[2] C.D.S. Brites, S. Balabhadra, L.D. Carlos, Lanthanide-Based Thermometers: At the Cutting-Edge of Luminescence Thermometry, Adv Opt Mater 7 (2019) 1801239. https://doi.org/10.1002/ADOM.201801239.

[3] L. Marciniak, K. Kniec, K. Elżbieciak-Piecka, K. Trejgis, J. Stefanska, M. Dramićanin, Luminescence thermometry with transition metal ions. A review, Coord Chem Rev 469 (2022). https://doi.org/10.1016/J.CCR.2022.214671.

[4] M.D. Dramićanin, Trends in luminescence thermometry, J Appl Phys 128 (2020). https://doi.org/10.1063/5.0014825.

[5] M. Aldén, A. Omrane, M. Richter, G. Särner, Thermographic phosphors for thermometry: A survey of combustion applications, Prog Energy Combust Sci 37 (2011) 422–461. https://doi.org/10.1016/J.PECS.2010.07.001.

[6] C.D.S. Brites, R. Marin, M. Suta, A.N. Carneiro Neto, E. Ximendes, D. Jaque, L.D. Carlos, Spotlight on Luminescence Thermometry: Basics, Challenges, and Cutting-Edge Applications, Advanced Materials 35 (2023) 2302749. https://doi.org/10.1002/ADMA.202302749.

[7] N. Fuhrmann, J. Brübach, A. Dreizler, Phosphor thermometry: A comparison of the luminescence lifetime and the intensity ratio approach, Proceedings of the Combustion Institute 34 (2013) 3611–3618. https://doi.org/10.1016/J.PROCI.2012.06.084.

[8] M. Fujiwara, Y. Shikano, Diamond quantum thermometry: from foundations to applications, Nanotechnology 32 (2021) 482002. https://doi.org/10.1088/1361-6528/AC1FB1.

[9] M. Aldén, A. Omrane, M. Richter, G. Särner, Thermographic phosphors for thermometry: A survey of combustion applications, Prog Energy Combust Sci 37 (2011) 422–461. https://doi.org/10.1016/J.PECS.2010.07.001.

[10] L. Marciniak, W. Piotrowski, M. Szalkowski, V. Kinzhybalo, M. Drozd, M. Dramicanin, A. Bednarkiewicz, Highly sensitive luminescence nanothermometry and thermal imaging facilitated by phase transition, Chemical Engineering Journal 427 (2022) 131941. https://doi.org/10.1016/J.CEJ.2021.131941.

[11] A. Javaid, M. Szymczak, M. Kubicka, V. Kinzhybalo, M. Drozd, D. Szymanski, L. Marciniak, Luminescent Platform for Thermal Sensing and Imaging Based on Structural Phase-Transition, Advanced Science (2025). https://doi.org/10.1002/ADVS.202508920.

[12] G. Sutton, S. Korniliou, A. Andreu, D. Wilson, Imaging Luminescence Thermometry to 750 °C for the Heat Treatment of Common Engineering Alloys and Comparison with Thermal Imaging, Int J Thermophys 43 (2022). https://doi.org/10.1007/S10765-021-02963-1.

[13] D. Avram, I. Porosnicu, A. Patrascu, C. Tiseanu, Real-Time Thermal Imaging based on the Simultaneous Rise and Decay Luminescence Lifetime Thermometry, Adv Photonics Res 3 (2022). https://doi.org/10.1002/ADPR.202100208.

[14] A.H. Khalid, K. Kontis, 2D surface thermal imaging using rise-time analysis from laser-induced luminescence phosphor thermometry, Meas Sci Technol 20 (2009). https://doi.org/10.1088/0957-0233/20/2/025305.

[15] Y. Wu, F. Li, Y. Wu, H. Wang, L. Gu, J. Zhang, Y. Qi, L. Meng, N. Kong, Y. Chai, Q. Hu, Z. Xing, W. Ren, F. Li, X. Zhu, Lanthanide luminescence nanothermometer with working wavelength beyond 1500 nm for cerebrovascular temperature imaging in vivo, Nature Communications 2024 15:1 15 (2024) 2341-. https://doi.org/10.1038/s41467-024-46727-5.